%% file: arxiv_main.tex
\documentclass[lettersize,journal]{IEEEtran}

\def\BibTeX{{\rm B\kern-.05em{\sc i\kern-.025em b}\kern-.08em
    T\kern-.1667em\lower.7ex\hbox{E}\kern-.125emX}}

\usepackage{rotating}
\usepackage{amssymb}
\usepackage{algorithm}
\usepackage{algpseudocode}
\usepackage{color}
\usepackage{epsf}
\usepackage{psfrag}
\usepackage{epsfig}
\usepackage{multicol}
\usepackage{multirow}
\usepackage{makecell}
\usepackage{amsfonts}
\usepackage{textcomp}
\usepackage{comment}
\usepackage{footnote}
\usepackage{tablefootnote}
\usepackage[flushleft]{threeparttable}
\usepackage{paralist}
\usepackage{mdwlist}
\usepackage{amssymb}
\usepackage{amsmath}
\usepackage{url}
\usepackage{verbatim}
\usepackage{alltt}
\usepackage{balance}
\usepackage{multirow}
\usepackage{epstopdf}
\usepackage{lipsum}
\usepackage{float}
\usepackage{booktabs}
\usepackage{slashbox}
\usepackage[table]{xcolor}
\usepackage{array}
\usepackage{boldline}
\usepackage{caption,setspace}
\usepackage{mathtools}
\usepackage[table]{xcolor}
\usepackage{csquotes}

\usepackage{dsfont}
\usepackage{xcolor}

\usepackage{amsthm}

\usepackage{stackengine}[2013-10-15]

\usepackage[T1]{fontenc}
\usepackage{frcursive}
\usepackage{calligra}
\usepackage{wedn}
\usepackage{bm}
\usepackage{aurical}
\usepackage{upgreek}

\definecolor{darkgreen}{RGB}{0,204,0}

\usepackage{cite}
\usepackage{adjustbox}

\usepackage{esint}
\usepackage{tipa}
\usepackage{soul}
\usepackage{tikz}
\usepackage{listings}
\usepackage{multicol}
\usepackage{longtable} 
\usepackage{enumitem}
\usepackage{mdframed}

\newcolumntype{?}{!{\vrule width 1pt}}
\newcolumntype{+}{!{\vrule width 1.25pt}}

\def\hlineb#1{%
\noalign{\ifnum0=`}\fi\hrule \@height #1 %
\futurelet\reserved@a\@xhline}

\usepackage{cancel}
\usepackage{subcaption}

\usepackage{pgfplots}
\pgfplotsset{compat=1.18} 
\usepackage{mathrsfs}
\usepackage{pifont}

\usepackage[most]{tcolorbox}
\tcbuselibrary{listingsutf8}

\newcolumntype{L}[1]{>{\raggedright\arraybackslash}m{#1}}
\newcolumntype{C}[1]{>{\centering\arraybackslash}m{#1}}

\begin{document}

\title{Modality Agreement- and Conflict-Aware Prototype Hypergraph Learning for Multimodal Intent Understanding}

\author{Mohnish Raj,
        Suraj Kumar, ~\IEEEmembership{Graduate Student Member, IEEE},
        Soumi Chattopadhayay, ~\IEEEmembership{Senior Member, IEEE},
        Chandranath Adak, ~\IEEEmembership{Senior Member, IEEE},
        Ayan Dutta
\thanks{
  Corresponding author: Soumi Chattopadhyay, email: soumi@iiti.ac.in.
  This work has been submitted to a prominent venue for possible publication. Copyright may be transferred without notice, after which this version may no longer be accessible.
}
}



\maketitle

\begin{abstract}
    Multimodal intent recognition requires understanding not only what textual, acoustic, and visual signals share, but also how they disagree. Such disagreement is frequently class-informative; for example, lexical positivity accompanied by incongruent vocal or facial behavior may indicate sarcasm or taunting, yet most fusion methods either encourage modality alignment or treat inconsistency as uncertainty to be suppressed. We propose MACH (Modality Agreement- and Conflict-aware prototype Hypergraph), a hierarchical prototype-hypergraph framework that represents multimodal agreement and conflict as distinct, recurring relational structures. MACH progressively composes unimodal representations into bimodal and trimodal abstractions. At each applicable level, modality-composition anchors activate sparse agreement prototype hypergraphs that capture reusable consensus patterns, while a separate conflict pathway maps cross-modal discrepancies to dedicated conflict prototype hypergraphs. The two pathways are combined through a feature-wise, sample-adaptive arbitration mechanism, enabling the model to preserve informative disagreement while suppressing incidental modality noise. A progressive optimization strategy stabilizes the interdependent hierarchy before joint agreement-conflict learning. Experiments on benchmark datasets demonstrate the effectiveness of the proposed formulation, while component and robustness analyses validate the distinct roles of hierarchical composition, prototype-mediated semantic refinement, and agreement-conflict arbitration.
\end{abstract}


\input{1_intro_new}
\input{2_related_Work}
\input{3_new_method}
\input{4_experiments}
\input{5_conclusion}
\bibliographystyle{ieeetr}
\bibliography{aaai2027}

\section*{Supplementary Appendix}
Appendices are provided at \url{https://github.com/csksuraj17/MACH}

\end{document}

%% file: 1_intro_new.tex
\section{Introduction}
Communicative intent is inherently multimodal: speakers convey meaning not only through lexical content, but also through prosody, facial expression, and other nonverbal cues~\cite{zhu2025survey,zhao2025deep}. Crucially, these modalities do not always provide consistent evidence. In many utterances, intent is revealed by how modalities reinforce, qualify, or contradict one another rather than by any modality in isolation. For instance, the same positive sentence may express appreciation when accompanied by congruent vocal and visual signals, but sarcasm or taunting when paired with an incongruent tone or expression. Conversely, intents such as requesting assistance may be characterized by coherent semantic and interpersonal cues across modalities. These examples suggest that multimodal intent recognition should model not only modality-specific content, but also recurring patterns of cross-modal agreement and conflict that distinguish communicative intents.

Despite substantial progress in multimodal intent recognition (MIR), existing methods predominantly model cross-modal interaction from two perspectives. The first seeks to reduce modality discrepancy by projecting textual, acoustic, and visual signals into a shared representation through cross-attention, tensor fusion, or modality-aware transformers~\cite{mult_2019_acl,magbert,MVCL_DAF_aaai2025,LiPLC25,GMoE-DCAA}. The second estimates modality reliability or confidence and uses disagreement to down-weight potentially unreliable signals during fusion~\cite{cdpr_icmr_2026,wcfmir_aaai26}. Although both paradigms improve multimodal representation learning, they assign only an implicit role to agreement and conflict: agreement is absorbed into the aligned representation, while conflict is primarily used as a sample-specific reliability cue. Consequently, they do not explicitly model recurring cross-modal agreement and conflict patterns shared across utterances that provide discriminative evidence for communicative intent.

We hypothesize that agreement and conflict are not merely sample-specific interaction relationships, but reusable higher-order interaction structures shared across semantically related utterances. While the lexical or visual content of two utterances may differ substantially, they can exhibit similar patterns of cross-modal agreement or disagreement that characterize the same communicative intent. Consequently, instead of modeling modality interactions independently for each utterance, MIR should learn structured interaction representations that capture and reuse these recurring interaction patterns as transferable semantic evidence.

Multimodal interaction is inherently hierarchical~\cite{hier_cvpr_26}. Intent emerges progressively as unimodal evidence is composed into bimodal interactions and ultimately holistic trimodal semantics. Consequently, structured interaction representations should also be learned progressively across successive modality-composition levels rather than in a single fusion step. To realize this idea, we employ prototype hypergraphs, which naturally capture higher-order relationships while enabling semantically related utterances to share reusable interaction structures across the hierarchy.

Motivated by these observations, we propose \textbf{MACH} (\textbf{M}ultimodal \textbf{A}greement- and \textbf{C}onflict-aware prototype \textbf{H}ypergraph), a hierarchical framework for \emph{Structured Agreement-Conflict Learning} in multimodal intent recognition. Rather than collapsing multimodal evidence into a single fused representation, MACH explicitly learns agreement and conflict as two complementary interaction spaces that capture reusable patterns of cross-modal semantics across successive modality-composition levels. Prototype hypergraphs provide a structured interaction memory that enables semantically related utterances to share higher-order interaction structures, while a sample-adaptive arbitration mechanism dynamically determines the relative contribution of agreement and conflict for intent prediction. Finally, a progressive optimization strategy aligns the learning process with the hierarchical dependency of interaction reasoning, enabling stable construction of increasingly abstract interaction representations. Our contributions are summarized as:

\begin{itemize}
    \item We introduce Structured Agreement-Conflict Learning for multimodal intent recognition from the perspective of reusable higher-order interaction representations, where agreement and conflict are explicitly modeled as structured relational patterns rather than implicit consequences of multimodal alignment.

    \item We propose \textbf{MACH}, a hierarchical prototype-hypergraph framework that progressively constructs reusable agreement and conflict interaction cues from unimodal to bimodal and trimodal interactions, together with adaptive agreement-conflict arbitration for intent prediction.

    \item Extensive experiments on three benchmark datasets demonstrate the effectiveness and robustness of MACH, with qualitative analyses showing that the learned interaction representations capture meaningful agreement and conflict patterns across diverse intent categories.
\end{itemize}

%% file: 2_related_Work.tex
\section{Related Work}
\textbf{Multimodal Intent Recognition (MIR):}
MIR aims to infer communicative intent by jointly reasoning over textual, acoustic, and visual signals. Early methods primarily relied on cross-modal attention and multimodal fusion to learn unified representations~\cite{mult_2019_acl,magbert,misa_MM_20}. More recent approaches improve representation quality through modality-specific alignment~\cite{TCLMAP_aaai_2024,MVCL_DAF_aaai2025}, contrastive learning~\cite{huang2024sdif}, and prompt- or instruction-based adaptation~\cite{hier_cvpr_26}. Despite increasingly discriminative embeddings, these methods ultimately optimize a single fused representation in which agreement and disagreement are implicitly absorbed, without explicitly modeling reusable interaction patterns.

\noindent
\textbf{Agreement-Conflict Modeling:}
Recognizing that modalities may provide complementary or conflicting evidence, recent studies have begun incorporating disagreement into multimodal reasoning. ECFMIR~\cite{wcfmir_aaai26} estimates modality confidence to suppress unreliable signals, while CDPR~\cite{cdpr_icmr_2026} and InMu-Net~\cite{InMuNet_2024_MM} treat disagreement as an auxiliary reasoning cue during multimodal fusion. Nevertheless, conflict is still treated primarily as a sample-specific reliability signal, without explicitly modeling agreement and conflict as complementary semantic spaces or learning transferable interaction patterns across semantically related utterances.

\noindent
\textbf{Prototype and Hypergraph Representation Learning:}
Prototype learning provides interpretable semantic representations by organizing samples around representative prototypes~\cite{PIP-NET_TMM25,huang2025mvcl}, while hypergraph neural networks naturally capture higher-order relationships beyond pairwise interactions~\cite{hypergraph_NN_aaai_2019}. Recent multimodal hypergraph methods, including GMoE-DCAA~\cite{GMoE-DCAA}, H2GF-Net~\cite{H2GF-Net}, HyperModal~\cite{al2025hypermodal}, and HyperGCL~\cite{saifuddin2025hypergcl}, leverage higher-order message passing to strengthen multimodal representations. However, prototypes are mainly used for class discrimination and hypergraphs for feature propagation, rather than representing reusable cross-modal interaction structures or their progressive evolution across modality-composition levels.


Overall, existing work advances multimodal representation learning, disagreement-aware fusion, prototype learning, and hypergraph reasoning along largely independent directions. However, none formulates multimodal intent recognition as learning structured agreement and conflict representations that are reusable across semantically related utterances and progressively constructed over hierarchical modality compositions. MACH addresses this gap through a unified Structured Agreement-Conflict Learning framework, where prototype-guided hypergraphs serve as the representation mechanism for reusable interaction structures rather than the primary contribution, fundamentally differing from fusion-based approaches that treat agreement and conflict as transient by-products of multimodal fusion.

%% file: 3_new_method.tex
\section{Our Method}\label{sec:method}

\begin{figure*}
    \centering
    \includegraphics[width=18cm, height=7.14cm]{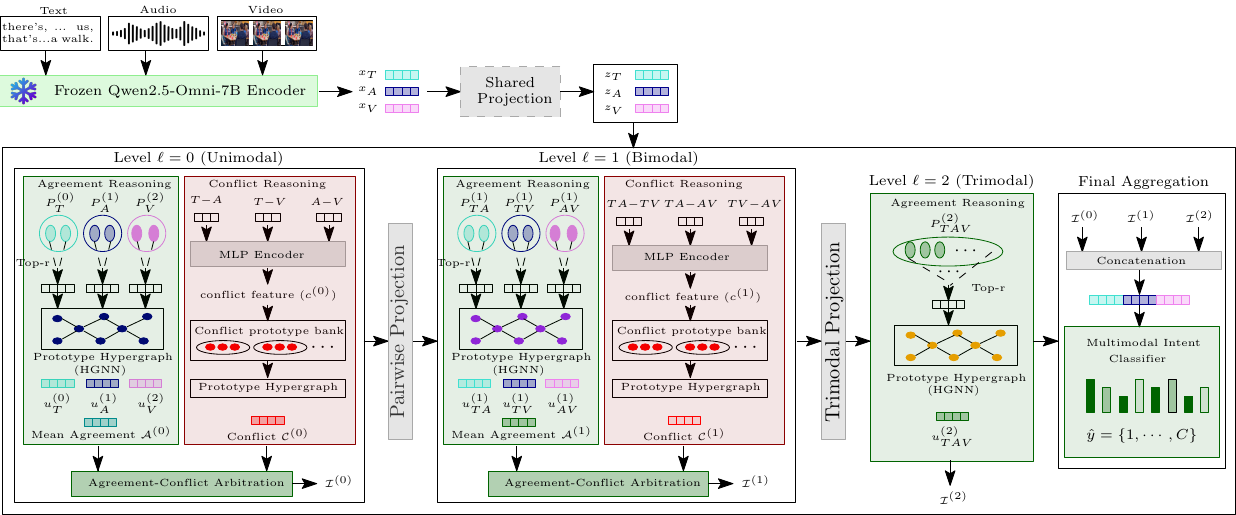}
    \caption{MACH full framework.}
    \label{fig:mach_arch}
\end{figure*}
\subsection{Problem Formulation}
Multimodal intent recognition aims to infer communicative intent $y \in \{1,\ldots,C\}$ from a synchronized utterance comprising text $T$, audio $A$, and video $V$. Let $\mathbf{x}_T \in \mathbb{R}^{d_T}$, $\mathbf{x}_A \in \mathbb{R}^{d_A}$, and $\mathbf{x}_V \in \mathbb{R}^{d_V}$ denote the corresponding utterance-level modality embeddings.
Conventional approaches formulate the task as learning a mapping
\(
f:(\mathbf{x}_T,\mathbf{x}_A,\mathbf{x}_V)\rightarrow y,
\) where reasoning is performed on a fused representation obtained by
modality alignment or aggregation.

However, communicative intent depends not only on individual modality semantics but also on \emph{how modalities interact}. Cross-modal disagreement can itself be highly discriminative (e.g., sarcastic text accompanied by incongruent vocal or facial cues), yet collapsing modalities into a single fused representation may obscure these interaction patterns.

Motivated by this observation, we reformulate multimodal intent recognition as learning \emph{structured cross-modal interaction representations} rather than fused feature representations. Specifically, we learn complementary agreement and conflict mappings,
\(\mathcal{A}=\Phi_{\mathrm{agr}}(\mathbf{x}_T,\mathbf{x}_A,\mathbf{x}_V), ~
\mathcal{C}=\Phi_{\mathrm{con}}(\mathbf{x}_T,\mathbf{x}_A,\mathbf{x}_V),
\)
where $\mathcal{A}$ and $\mathcal{C}$ capture reusable agreement and conflict structures, respectively. The final prediction is obtained as \(\hat{y}=g(\mathcal{A},\mathcal{C}),\) where $g(\cdot)$ jointly reasons over the two complementary interaction representations.

\subsection{Structured Agreement-Conflict Learning}
Let $\mathbf{X}=\{\mathbf{x}_T,\mathbf{x}_A,\mathbf{x}_V\}$ denote the synchronized textual, acoustic, and visual representations of an utterance, where $\mathbf{x}_m\in\mathbb{R}^{d_m}$ for modality $m\in\{T,A,V\}$. Conventional multimodal intent recognition learns a feature mapping
\( F:\mathbf{X} \rightarrow \mathbf{z}, \)
followed by a classifier
\(\hat{y} = g(\mathbf{z}),\)
implicitly assuming that a single fused representation $\mathbf{z}$ sufficiently captures multimodal semantics.

We instead formulate intent recognition as learning \emph{interaction representations}. Specifically, we define an interaction mapping \(\Phi:\mathbf{X}\rightarrow\mathcal{I},\) where $\mathcal{I}$ denotes the latent interaction space. We decompose
\(\Phi=(\Phi_{\mathrm{agr}},\Phi_{\mathrm{con}}),\) to obtain complementary agreement and conflict representations,
\(\mathcal{A}=\Phi_{\mathrm{agr}}(\mathbf{X}), ~ \mathcal{C}=\Phi_{\mathrm{con}}(\mathbf{X}),\) where $\mathcal{A}$ captures cross-modal semantic agreement and $\mathcal{C}$ captures informative semantic conflict. Rather than absorbing disagreement into feature fusion or treating it as modality unreliability, we model agreement and conflict as complementary latent variables that jointly characterize communicative intent. The prediction is then obtained as
\(\hat{y}=g(\mathcal{A},\mathcal{C}).\)
We hypothesize that semantically related utterances share transferable agreement and conflict structures despite differences in their lexical, acoustic, and visual realizations. MACH realizes this formulation through progressive modality composition, where agreement captures consistent semantic evidence and conflict preserves discriminative semantic inconsistencies. By explicitly separating these complementary interaction spaces, MACH models reusable interaction structures instead of a single fused embedding.
\subsection{Progressive Interaction Reasoning}
The interaction formulation specifies \emph{what} should be learned, namely the agreement and conflict representations, but not \emph{how} they should be constructed. We formulate interaction reasoning as a progressive composition process, motivated by the observation that multimodal semantics emerge through increasingly richer modality compositions.

Let \(\mathcal{S}^{(0)}=\{T,A,V\}; ~\mathcal{S}^{(1)}=\{TA,TV,AV\}; ~\mathcal{S}^{(2)}=\{TAV\}, \)
denote the modality-composition sets corresponding to the unimodal, pairwise, and holistic reasoning levels, respectively. The interaction representation constructed over $\mathcal{S}^{(\ell)}$ is denoted by $\mathcal{I}^{(\ell)}$. For the unimodal and pairwise levels,
\(\mathcal{I}^{(\ell)} = f_1\left(\mathcal{A}^{(\ell)}, \mathcal{C}^{(\ell)}\right),~\ell\in\{0,1\},\)
where $\mathcal{A}^{(\ell)}$ and $\mathcal{C}^{(\ell)}$ are the agreement and conflict representations. Since $\mathcal{S}^{(2)}$ contains a single trimodal composition, no additional conflict interaction can be defined, yielding
\(\mathcal{I}^{(2)}=f_2(\mathcal{A}^{(2)}).\)

The three reasoning levels operate on
\(\mathbf{X}^{(0)}=\{\mathbf{x}_T,\mathbf{x}_A,\mathbf{x}_V\},\)
their pairwise compositions, and the complete trimodal composition, respectively. Interaction reasoning proceeds sequentially,
\(\mathcal{I}^{(0)} \rightarrow \mathcal{I}^{(1)} \rightarrow \mathcal{I}^{(2)},\)
such that higher-level representations progressively enrich the semantic evidence learned at preceding levels. The final intent prediction jointly exploits representations from all reasoning levels,
\(\hat{y} = g\left(\mathcal{I}^{(0)}, \mathcal{I}^{(1)}, \mathcal{I}^{(2)}\right).\)
MACH realizes this formulation by progressively constructing the interaction representations across the hierarchy.
\subsection{MACH Overview}
We realize the proposed structured agreement-conflict learning formulation using \textbf{MACH} (\textbf{M}odality \textbf{A}greement-and \textbf{C}onflict-aware Prototype \textbf{H}ypergraphs), a hierarchical interaction reasoning framework operating over the modality-composition hierarchy defined by $\{\mathcal{S}^{(\ell)}\}_{\ell=0}^{2}$. At each reasoning level $\ell$, MACH realizes the interaction representation $\mathcal{I}^{(\ell)}$ by constructing the agreement and conflict representations, $\mathcal{A}^{(\ell)}$ and $\mathcal{C}^{(\ell)}$, through prototype-guided hypergraph reasoning and explicit conflict modeling, respectively. MACH progressively constructs higher-order cross-modal interaction representations through hierarchical modality composition, while prototype-guided hypergraphs refine each modality-composition anchor using reusable semantic prototypes.

\subsection{Structured Agreement Representation}
For each reasoning level $\ell$, MACH realizes the agreement representation $\mathcal{A}^{(\ell)}$ through three stages: interaction anchor generation, prototype-guided interaction modeling, and agreement propagation.

\emph{Interaction Anchors Generation:}
Modality-specific embeddings are extracted using 
a multimodal encoder.
After fine-tuning, the encoders are frozen, and each modality embedding is projected into a common $d$-dimensional interaction space:
\begin{equation}\scriptsize
    \mathbf{z}^{(0)}_{m} = \mathrm{LN}\left(W_{2}\sigma(W_{1}\mathbf{x}_{m})\right),\qquad m\in\{T,A,V\},
    \label{eq:encoder}
\end{equation}
where $\mathbf{x}_{m}\in\mathbb{R}^{d_m}$ is the encoder output for modality $m$, $W_{1}$ and $W_{2}$ are learnable projection matrices, $\sigma(\cdot)$ denotes GELU, and $\mathbf{z}^{(0)}_{m}\in\mathbb{R}^{d}$ is the corresponding unimodal interaction anchor.

Higher-level anchors are constructed recursively from the updated agreement representations of the preceding level. Let $\mathbf{u}^{(\ell)}_{s}\in\mathbb{R}^{d}$ denote the post-hypergraph representation of interaction unit $s\in\mathcal{S}^{(\ell)}$. The bimodal anchors are
\begin{equation}\scriptsize
\mathbf{z}^{(1)}_{ij} = \psi^{(1)}_{ij}\left(\left[\mathbf{u}^{(0)}_{i};\mathbf{u}^{(0)}_{j}\right]\right),\qquad ij\in\{TA,TV,AV\},
\label{eq:bimodal_anchor}
\end{equation}
and the trimodal anchor is 
\begin{equation}\scriptsize
\mathbf{z}^{(2)}_{TAV} = \psi^{(2)}\left(\left[\mathbf{u}^{(1)}_{TA};\mathbf{u}^{(1)}_{TV};\mathbf{u}^{(1)}_{AV}\right]\right),
\label{eq:trimodal_anchor}
\end{equation}
where $[\cdot\,;\cdot]$ denotes concatenation and $\psi^{(1)}_{ij}$ and $\psi^{(2)}$ are learnable projection functions mapping their inputs to $\mathbb{R}^{d}$. Hence, the level-wise anchor sets are
\begin{equation}\scriptsize
\mathcal{Z}^{(0)} = \{\mathbf{z}^{(0)}_{T},\mathbf{z}^{(0)}_{A},\mathbf{z}^{(0)}_{V}\}; ~~
\mathcal{Z}^{(1)} = \{\mathbf{z}^{(1)}_{TA},\mathbf{z}^{(1)}_{TV},\mathbf{z}^{(1)}_{AV}\}; ~~
\mathcal{Z}^{(2)} = \{\mathbf{z}^{(2)}_{TAV}\}. 
\end{equation}

\emph{Prototype-Guided Interaction Modeling:}
Agreement interactions exhibit recurring semantic structures across semantically related utterances. Instead of learning these structures independently for every sample, MACH maintains a compact set of learnable agreement prototypes that serve as reusable semantic memories for each interaction unit.

At reasoning level $\ell$, let
\(\mathcal{Z}^{(\ell)} = \{\mathbf{z}^{(\ell)}_{s}\mid s\in\mathcal{S}^{(\ell)}\},\)
denote the set of interaction anchors, where
$|\mathcal{S}^{(\ell)}|=\Lambda$
($\Lambda=3$ for $\ell\in\{0,1\}$ and $\Lambda=1$ for $\ell=2$).
Each interaction unit
$s\in\mathcal{S}^{(\ell)}$
is associated with an agreement prototype bank
\(P^{(\ell)}_{s} = \left\{\mathbf{p}^{(\ell)}_{s,k}\right\}_{k=1}^{K},~ P^{(\ell)}_{s}\in\mathbb{R}^{K\times d},\)
where each prototype
$\mathbf{p}^{(\ell)}_{s,k}\in\mathbb{R}^{d}$
represents a recurring agreement pattern learned from the training data. 
For every interaction anchor $\mathbf{z}^{(\ell)}_{s}$, its similarity to the prototype bank is computed using cosine similarity
\(\gamma^{(\ell)}_{s,k} = cosine(\mathbf{z}^{(\ell)}_{s}, \mathbf{p}^{(\ell)}_{s,k})\), for $k=1,\ldots,K$.
%
Only the $r$ highest-scoring prototypes are retained,
\(\mathcal{N}_{r}(s) = \operatorname{Top}\text{-}r \left(\{\gamma^{(\ell)}_{s,k}\}_{k=1}^{K}\right).\)

These sparse anchor-prototype assignments define the prototype-guided hypergraph
\(\mathcal G^{(\ell)} = (\mathcal V^{(\ell)}, \mathcal E^{(\ell)}).\) Its vertex set contains both the interaction-anchor nodes and their associated prototype nodes:
\(\mathcal V^{(\ell)} = \mathcal V_{\mathrm{anc}}^{(\ell)} \cup \mathcal V_{\mathrm{pro}}^{(\ell)},\) where
\(\mathcal V_{\mathrm{anc}}^{(\ell)} = \{v_s\mid s\in\mathcal S^{(\ell)}\}, \mathcal V_{\mathrm{pro}}^{(\ell)} = \{v_{s,k}^{p} \mid s\in\mathcal S^{(\ell)},\, k=1,\ldots,K\}. \)
The initial features of $v_s$ and $v_{s,k}^{p}$ are respectively $\mathbf z_s^{(\ell)}$ and $\mathbf p_{s,k}^{(\ell)}$.
For each interaction unit $s$, one hyperedge is constructed by connecting its anchor to the top-$r$ nearest prototypes from its unit-specific prototype bank:
\(
e_s^{(\ell)} = \{v_s\} \cup \left\{v_{s,k}^{p} \mid k\in\mathcal N_r(s)\right\}.\)
Thus,
\(\mathcal E^{(\ell)} = \{e_s^{(\ell)}\mid s\in\mathcal S^{(\ell)}\}, |\mathcal E^{(\ell)}|=\Lambda.\)
Let \(N^{(\ell)} = \Lambda+\sum_{s\in\mathcal S^{(\ell)}}K \)
denote the total number of anchor and prototype nodes. The incidence matrix
\(H^{(\ell)} \in \{0,1\}^{N^{(\ell)}\times \Lambda}\) is defined as
\[
H^{(\ell)}(v,e_s)
=
\begin{cases}
1,
&
v=v_s,
\\
1,
&
v=v_{s,k}^{p}
\text{ for some }k\in\mathcal N_r(s),
\\
0,
&
\text{otherwise}.
\end{cases}
\]
Hence, each hyperedge jointly represents an interaction anchor and the subset of prototypes that best characterize its agreement semantics.
%
%
By restricting each interaction anchor to its nearest prototypes, MACH preserves localized agreement semantics while preventing message propagation through weak or semantically unrelated prototype associations, resulting in sparse, interpretable, and computationally efficient higher-order interaction graphs.


\emph{Agreement Propagation:}
The prototype-guided hypergraph propagates agreement information over interaction-anchor and prototype nodes, where hyperedges connect each interaction anchor to its top-$r$ activated prototypes. Let \(X^{(\ell)}\in\mathbb{R}^{N^{(\ell)}\times d}\) denote the node feature matrix. Agreement propagation is performed using the normalized HGNN
\begin{equation}\scriptsize
U^{(\ell)} = \phi\left((D_v^{(\ell)})^{-1/2}H^{(\ell)}W_e^{(\ell)}(D_e^{(\ell)})^{-1}(H^{(\ell)})^\top(D_v^{(\ell)})^{-1/2}X^{(\ell)}\Theta^{(\ell)}\right),
\label{eq:hgnn}
\end{equation}
where $H^{(\ell)}$ is the incidence matrix, $D_v^{(\ell)}$ and $D_e^{(\ell)}$ are the node and hyperedge degree matrices, $W_e^{(\ell)}$ is the hyperedge-weight matrix, $\Theta^{(\ell)}$ is a learnable projection, and $\phi(\cdot)$ is the activation function. Residual connections and LayerNorm are applied after propagation. The updated interaction-anchor representations $\{\mathbf{u}_s^{(\ell)}\}_{s\in\mathcal S^{(\ell)}}$ are retained, and the level-wise agreement representation is computed as
\(\mathcal A^{(\ell)} = \frac{1}{\Lambda}\sum_{s\in\mathcal S^{(\ell)}}\mathbf{u}_s^{(\ell)}.\)
\subsection{Structured Conflict Representation}
Hypergraph propagation first reinforces shared semantic evidence. The remaining differences among the updated interaction anchors capture complementary disagreement. Accordingly, conflict is computed from pairwise absolute differences between updated interaction anchors.
For a hierarchy level with $\Lambda\geq2$ groups, the conflict feature is computed from all pairwise interaction-anchor differences,
\begin{equation}\scriptsize
\label{eq:conflict}
\mathbf{c}^{(\ell)} = \rho^{(\ell)}\left(\left[\left|\mathbf{u}^{(\ell)}_{s_1}-\mathbf{u}^{(\ell)}_{s_2}\right|\right]_{\substack{s_1,s_2\in\mathcal S^{(\ell)}\\s_1<s_2}}\right),
\end{equation}
where $\rho^{(\ell)}:\mathbb{R}^{\binom{\Lambda}{2}d}\rightarrow\mathbb{R}^{d}$ is a two-layer fully connected network. At the trimodal level ($\Lambda=1$), no pairwise disagreement exists; consequently, only the agreement branch is retained. 
The aggregated conflict feature $\mathbf{c}^{(\ell)}$ is then treated as a single conflict anchor and connected to its top-$r_c$ prototypes from a level-specific conflict bank $\mathcal Q^{(\ell)}=\{\mathbf q_k^{(\ell)}\}_{k=1}^{K_c}$. HGNN propagation over this anchor-centered prototype hypergraph yields the updated conflict representation $\mathcal C^{(\ell)}$. Maintaining a separate prototype bank enables agreement and conflict prototypes to capture complementary consensus and disagreement patterns without interference.
\subsection{Agreement-Conflict Arbitration}
Agreement and conflict provide complementary evidence with sample- and feature-dependent importance. We therefore perform adaptive feature-wise arbitration.
\begin{equation}
\scriptsize
\begin{aligned}
\boldsymbol{\omega}^{(\ell)} &= \sigma\left(W_{\omega}^{(\ell)}\left[\mathcal A^{(\ell)}; \mathcal{C}^{(\ell)}\right] + \mathbf b_{\omega}^{(\ell)} \right),\\
\mathcal{I}^{(\ell)} &= \left(1-\boldsymbol{\omega}^{(\ell)}\right) \odot \mathcal A^{(\ell)} + \boldsymbol{\omega}^{(\ell)} \odot \mathcal{C}^{(\ell)},
\end{aligned}
\label{eq:gate}
\end{equation}
where $\boldsymbol{\omega}\in(0,1)^d$ denotes the feature-wise arbitration weight, $\sigma(\cdot)$ is the sigmoid function, and $\odot$ denotes element-wise multiplication. The arbitration adaptively balances agreement and conflict evidence for each feature according to the input sample.

\subsection{Progressive Interaction Composition}
The hierarchy is constructed recursively using Eqs.~\eqref{eq:bimodal_anchor}-\eqref{eq:trimodal_anchor}. Each level produces agreement and conflict representations, which are independently processed and fused. The final representation concatenates the outputs from all hierarchy levels before classification.
Since the final level contains a single interaction group ($\Lambda=1$), only the agreement representation is retained. The final prediction is obtained from
\begin{equation}
\scriptsize
\mathcal{I} = \left[\mathcal{I}^{(0)}; \mathcal{I}^{(1)}; \mathcal{I}^{(2)}\right];\quad
\hat{\mathbf y} = \mathrm{softmax}\left(\mathrm{FFN}(\mathcal{I})\right),
\label{eq:classify}
\end{equation}
where the FFN consists of two fully connected layers with GELU activation, LayerNorm, and dropout.
\subsection{Learning Objectives}
MACH is optimized using four complementary objectives.

\emph{Hierarchical alignment:} To preserve semantic consistency across hierarchy levels, each interaction anchor is aligned with its updated representation using the InfoNCE objective,
\begin{equation}
\label{eq:hier}
\scriptsize
\mathcal{L}_{\mathrm{hier}} = \sum_{\ell} \sum_{s\in\mathcal S^{(\ell)}} - \log\frac{\exp\left(\mathrm{sim}\left(\mathbf z_s^{(\ell)}, \mathbf u_s^{(\ell)}\right)/\tau\right)}{\sum_j \exp\left(\mathrm{sim}\left(\mathbf z_s^{(\ell)}, \mathbf u_{s}^{(\ell,j)}\right)/\tau\right)}.
\end{equation}
where samples from other instances in the mini-batch serve as negatives.

\emph{Prototype regularization:}
Prototype learning jointly optimizes (i) prototype relevance, (ii) prototype coverage, (iii) prototype diversity, and (iv) assignment entropy regularization, as defined in Eq.~\eqref{eq:proto}. 

\begin{equation}
\label{eq:proto}
\scriptsize
\begin{aligned}
\mathcal L_{\mathrm{proto}}^{(\ell)}
=&{}
-\frac{1}{B\Lambda}
\sum_{b=1}^{B}
\sum_{s\in\mathcal S^{(\ell)}}
\max_{k}
\cos\left(
\mathbf z_{b,s}^{(\ell)},
\mathbf p_{s,k}^{(\ell)}
\right)
\\
&+
\frac{1}{\Lambda K}
\sum_{s\in\mathcal S^{(\ell)}}
\sum_{k=1}^{K}
\left[
1-
\max_{1\leq b\leq B}
\cos\left(
\mathbf z_{b,s}^{(\ell)},
\mathbf p_{s,k}^{(\ell)}
\right)
\right]
\\
&+
\frac{1}{\Lambda K(K-1)}
\sum_{s\in\mathcal S^{(\ell)}}
\sum_{i\neq j}
\cos^2\left(
\mathbf p_{s,i}^{(\ell)},
\mathbf p_{s,j}^{(\ell)}
\right)
\\
&-
\frac{\lambda_{\mathrm{ent}}}{B\Lambda}
\sum_{b=1}^{B}
\sum_{s\in\mathcal S^{(\ell)}}
\sum_{k=1}^{K}
\alpha_{b,s,k}^{(\ell)}
\log\alpha_{b,s,k}^{(\ell)} 
\end{aligned}
\end{equation}


\noindent 
where $B$ denotes the mini-batch size, $\mathbf z_{b,s}^{(\ell)}$ is the interaction anchor corresponding to the $s$-th interaction unit of the $b$-th training sample at hierarchy level $\ell$, and $\mathbf p_{s,k}^{(\ell)}$ is the $k$-th prototype in the prototype bank associated with interaction unit $s$. The coefficient $\lambda_{\mathrm{ent}}$ controls the strength of the entropy regularization, while $\tau$ is the temperature parameter of the prototype assignment distribution. The soft assignment weight is defined as 
\(
\alpha_{b,s,k}^{(\ell)}
=
\frac{
\exp\left(
\cos(
\mathbf z_{b,s}^{(\ell)},
\mathbf p_{s,k}^{(\ell)}
)/\tau
\right)
}{
\sum_{t=1}^{K}
\exp\left(
\cos(
\mathbf z_{b,s}^{(\ell)},
\mathbf p_{s,t}^{(\ell)}
)/\tau
\right)
}.
\)
The agreement and conflict prototype banks are optimized independently using Eq.~\eqref{eq:proto}, yielding $\mathcal{L}_{\text{agr}}$ and $\mathcal{L}_{\text{con}} = \mathcal{L}_{\mathrm{proto}}^{(\ell)}
\left(
\{\mathbf{c}_{b}^{(\ell)}\}_{b=1}^{B},
\mathcal{Q}^{(\ell)}
\right)$, respectively. 

\emph{Conflict contrastive loss:}
Residual conflict representations are regularized using the supervised contrastive loss \cite{cont_leanring_nips_2020},
\(\mathcal L_{\mathrm{sup-con}},\) encouraging samples with the same intent label to form compact conflict representations.

\emph{Classification loss:}
The classifier is trained using the cross-entropy loss with label smoothing, \(\mathcal L_{\mathrm{cls}}.\)

\emph{Overall loss:} The overall objective is: 
\begin{equation}
\scriptsize
\mathcal L = \lambda_{\mathrm{cls}}\mathcal L_{\mathrm{cls}} + \lambda_1\mathcal L_{\mathrm{hier}} + \lambda_2\mathcal L_{\mathrm{sup-con}} + (\lambda_3\mathcal L_{\mathrm{agr}} + \lambda_4\mathcal L_{\mathrm{con}}).
\label{eq:total}
\end{equation}

\noindent 
with $\lambda_1$ decayed across training as the hierarchy stabilizes, and $\lambda_{\text{cls}}$ linearly warmed up over the first epochs of any stage in which the classifier is newly introduced, so that classification gradients do not dominate before the classification head has moved off its random initialization.

\subsection{Progressive Optimization}
MACH is trained using a six-phase curriculum that follows the dependency of the interaction hierarchy. Phases~1-3 progressively optimize the agreement branch from $\ell=0$ to $\ell=2$, introducing each hierarchy level only after the preceding level has stabilized. The classifier and agreement-conflict arbitration are initialized after the complete agreement hierarchy is established. Phase~4 optimizes the conflict branch while freezing the agreement hierarchy. Finally, Phases~5-6 jointly optimize the complete model by first learning the agreement-conflict arbitration and then performing end-to-end fine-tuning. During progressive training, parameters inherited from earlier phases are updated using a smaller learning rate than newly introduced parameters.


%% file: 4_experiments.tex
\section{Experiments}\label{sec:experiments}


\input{tables/dataset}
\input{tables/sota_table}

\paragraph{Dataset Details.}
We evaluate MACH on three public multimodal benchmarks: MIntRec~\cite{mintrec} and MIntRec2.0~\cite{mintrec2} for multimodal intent recognition, and MELD-DA~\cite{meld2019_ACL} for multimodal emotion recognition. Together, these datasets cover diverse conversational scenarios by jointly modeling textual, acoustic, and visual modalities, providing a comprehensive evaluation of multimodal interaction understanding. Dataset statistics are summarized in Table~\ref{tab:dataset}.

\noindent
\paragraph{Experimental Settings.}
We used the given train/validation/test splits of MIntRec, MIntRec2.0, and MELD-DA for fair comparison with prior work. Qwen2.5-Omni-7B was adapted using QLoRA to extract multimodal representations. MACH was trained using the proposed six-phase curriculum (batch size 32), with early stopping (patience = 5) applied only in the final phase. 
All results are reported on the test set and averaged over five runs with seeds 0-4. Additional implementation details are given in the supplementary file.

\paragraph{Evaluation Metrics.}
We evaluate performance using accuracy (Acc), weighted precision (WP), weighted F1-score (WF1), and macro F1-score (F1). Weighted metrics are included to account for the class imbalance problem.

\subsection{Main Results and Comparative Study} Table~\ref{tab:results} compares MACH with representative multimodal fusion, confidence-aware reasoning, graph-based, and recent multimodal intent recognition methods on three benchmark datasets. Overall, MACH consistently achieves the best performance across all reported metrics, demonstrating the effectiveness and generalizability of the proposed framework.

Compared with conventional and recent multimodal fusion approaches (e.g., MulT~\cite{mult_2019_acl}, MAG-BERT~\cite{magbert}, MISA~\cite{misa_MM_20}, TCL-MAP~\cite{TCLMAP_aaai_2024}, and MVCL-DAF~\cite{MVCL_DAF_aaai2025}), MACH consistently delivers superior results on both MIntRec and MIntRec2.0. Likewise, it outperforms confidence-aware reasoning methods such as ECFMIR~\cite{wcfmir_aaai26} and CDPR~\cite{cdpr_icmr_2026}, indicating that explicitly modeling reusable agreement and conflict interaction representations is more effective than treating disagreement solely as a reliability cue. MACH also consistently surpasses 
hypergraph-based methods, including 
HyperModal~\cite{al2025hypermodal}, and HyperGCL~\cite{saifuddin2025hypergcl}, demonstrating the advantage of modeling higher-order multimodal interaction structures through prototype-guided hypergraphs.

Among all baselines, HIER provides the strongest overall performance by progressively modeling hierarchical multimodal semantics. Nevertheless, MACH consistently exceeds HIER across both intent benchmarks, improving Accuracy by $0.99\%$ on MIntRec and $1.52\%$ on MIntRec2.0, while also achieving higher weighted and macro F1-scores. Class-wise results (included in the supplementary) further confirm these improvements across intent categories. These results suggest that hierarchical composition alone is insufficient; explicitly modeling reusable agreement and conflict interaction representations provides complementary information.

The proposed framework also generalizes well to the conversational MELD-DA benchmark, achieving competitive performance on average across all metrics. 
Despite the greater ambiguity and contextual variability of conversational interactions, MACH maintains consistent improvements, demonstrating that the learned agreement-conflict interaction representations generalize effectively across diverse multimodal intent recognition scenarios.

\input{tables/ablation}

\subsection{Ablation Study}
To quantify the contribution of each component, we systematically remove individual modules from MACH and report the results in Table~\ref{tab:ablation}.

We first evaluate the frozen Qwen2.5-Omni-7B encoder by directly performing intent classification on its multimodal representations. Although it provides a strong baseline, MACH consistently achieves better performance on both datasets, showing that pretrained multimodal embeddings alone are insufficient without explicit agreement-conflict interaction modeling.

Next, we evaluate the proposed prototype-guided hypergraph. Replacing it with direct feature aggregation consistently reduces performance, confirming that higher-order interaction modeling produces more discriminative agreement representations than direct multimodal aggregation.

We further analyze the contribution of different modalities. Among the unimodal settings, text consistently performs best, while the visual modality alone contributes little. Combining text with audio yields the strongest modality pair, whereas pairs involving the visual modality achieve smaller gains, motivating the need for structured multimodal interaction reasoning rather than uniform modality fusion.

We then examine the hierarchical interaction architecture. Removing the hierarchy consistently degrades performance, demonstrating the benefit of progressively refining interactions from unimodal to bimodal and finally trimodal representations. Moreover, no individual level (Only Level-0/1/2) matches the complete hierarchy, indicating that different interaction levels capture complementary semantic information. The larger degradation after removing the bimodal level further highlights the importance of pairwise interactions.

The next group evaluates prototype learning. Removing the agreement prototypes causes one of the largest performance drops, emphasizing their role in learning reliable consensus patterns. In contrast, removing the conflict prototypes results in only a modest decline, suggesting that conflict representations remain informative even without prototype guidance but become more structured when dedicated conflict prototypes are introduced. Removing all prototypes affects MIntRec2.0 more noticeably, indicating that prototype-guided abstraction becomes increasingly beneficial for more complex intent spaces.

Finally, we validate the complementary roles of the agreement and conflict branches. Using either branch alone or removing either branch consistently underperforms the complete model, demonstrating that agreement captures shared semantic evidence while conflict preserves informative modality discrepancies. Their adaptive integration therefore produces more robust interaction representations, leading to the best overall performance.
\input{tables/loss_ablation}
\subsection{Loss Analysis}
Table~\ref{tab:loss_ablation} quantifies the contribution of each optimization objective. Removing the diversity loss causes the largest degradation on MIntRec ($3.46\%$ Accuracy and $3.63\%$ F1), confirming that encouraging diverse prototypes is fundamental for capturing complementary intent semantics rather than redundant representations. Eliminating the supervised contrastive loss results in a relatively smaller decline, suggesting that it primarily refines the learned feature space by improving intra-class compactness and inter-class discrimination. In contrast, the hierarchical loss has the greatest impact on MIntRec2.0, where its removal reduces Accuracy and F1 by $2.91\%$ and $2.85\%$, respectively. This indicates that enforcing consistency across hierarchical agreement representations becomes increasingly important in the more challenging and fine-grained benchmark. Finally, the full model consistently outperforms the classification-only variant on both datasets, demonstrating that the three objectives are complementary: diversity promotes informative prototypes, contrastive learning improves discriminability, and hierarchical supervision preserves semantic consistency across representation levels.

Further, we include supplementary material that provides additional experiments, including detailed implementation settings, class-wise performance analysis, hyperparameter sensitivity analysis, qualitative analysis, failure case analysis, feature representation visualization, detailed baseline descriptions, computational complexity analysis, and statistical significance tests.




%% file: tables/dataset.tex
\begin{table}[!b]
    \centering
    \caption{Evaluation dataset statistics.}
    \label{tab:dataset}
    \small
    \setlength{\tabcolsep}{2pt}
    \resizebox{\columnwidth}{!}{
        \begin{tabular}{lccc}
            \toprule
            \textbf{Attribute} & \textbf{MIntRec} & \textbf{MIntRec2.0} & \textbf{MELD-DA} \\
            \midrule
            Classes & 20 & 30 & 12 \\
            Video clips & 2,224 & 15,040 & 9988 \\
            Avg. utterance length (words) & 7.04 & 7.00 & 8.10 \\
            Avg. video duration (s) & 2.38 & 3.00 & 3.59 \\
            \bottomrule
        \end{tabular}
        }
\end{table}

%% file: tables/sota_table.tex
\begin{table*}[t]
    \centering
    \caption{Comparative study of MACH. Best and second-best results are shown in \textbf{bold} and \underline{underlined}, respectively.}
    \label{tab:results}
    \small
    \setlength{\tabcolsep}{5pt}
    \begin{tabular}{lcccccccccccc}
    \toprule
    \multirow{2}{*}{\textbf{Method}} &
    \multicolumn{4}{c}{\textbf{MIntRec}} &
    \multicolumn{4}{c}{\textbf{MIntRec2.0}} &
    \multicolumn{4}{c}{\textbf{MELD-DA}} \\
    \cmidrule(lr){2-5}
    \cmidrule(lr){6-9}
    \cmidrule(l){10-13}
    & \textbf{Acc} & \textbf{WP} & \textbf{WF1} & \textbf{F1}
    & \textbf{Acc} & \textbf{WP} & \textbf{WF1} & \textbf{F1}
    & \textbf{Acc} & \textbf{WP} & \textbf{WF1} & \textbf{F1} \\
    
    \midrule
    MISA
    & 72.29 & 73.48 & 72.38 & 69.32
    & 55.16 & 57.06 & 55.05 & 49.51
    & 60.86 & 59.55 & 58.80 & 49.45 \\
    
    MuIT
    & 72.52 & 72.85 & 72.31 & 69.25
    & 56.95 & 54.49 & 54.26 & 54.26
    & 59.99 & 59.39 & 58.67 & 50.69 \\
    
    MAG-BERT
    & 72.65 & 72.53 & 72.16 & 68.64
    & 55.87 & 53.71 & 52.58 & 52.58
    & 61.08 & 59.60 & 59.59 & 50.02 \\
    
    TCL-MAP
    & 73.17 & 72.97 & 72.66 & 68.92
    & 58.24 & 57.55 & 57.24 & 52.25
    & 61.63 & 60.10 & 59.74 & 50.25 \\
    
    
    MVCL-DAF
    & 73.63 & 74.31 & 73.57 & 70.41
    & 59.64 & 58.57 & 58.67 & 53.41
    & 60.78 & 59.83 & 59.16 & 48.88 \\
    
    \midrule
    
    ECFMIR
    & 74.38 & 75.05 & 74.51 & 70.83
    & 56.42 &58.44 & 55.82 & 50.02
    & 57.61 & 55.34 & 54.41 & 43.00 \\
    
    CDPR
    & 75.15 & 75.37 & 74.91 & 71.04
    & 60.82 & 60.23 & 59.54 & 53.86
    & 61.61 & 60.02 & \underline{60.49} & 51.30 \\
    
    \midrule
    
    
    HyperGCL
    & 79.11 & 79.17 & 78.74 & 76.06
    & 63.01 & 61.50 & 61.56 & 55.68
    & 60.06 & 55.29 & 56.73 & 41.58 \\
    
    HyperModal
    & 79.32 & 80.11 & 79.41 & \underline{77.27}
    & \underline{64.28} & \underline{65.53} & \underline{64.62} & \underline{60.64}
    & 57.11 & 58.35 & 56.75 & 50.98 \\
    
    \midrule
    
    
    HIER
    & \underline{80.00} & \underline{80.67} & \underline{79.59} & 76.91
    & 64.15 & 64.17 & 63.79 & 60.31
    & \underline{61.95} & \underline{60.44} & 60.38 & \textbf{54.80} \\
    
    \midrule
    
    \rowcolor{gray!12}
    \textbf{MACH (Ours)}
    & \makecell{\textbf{80.99}\\{\scriptsize$\pm$0.50}}
    & \makecell{\textbf{81.23}\\{\scriptsize$\pm$0.62}}
    & \makecell{\textbf{80.79}\\{\scriptsize$\pm$0.48}}
    & \makecell{\textbf{78.41}\\{\scriptsize$\pm$0.91}}
    & \makecell{\textbf{65.67}\\{\scriptsize$\pm$0.24}}
    & \makecell{\textbf{65.56}\\{\scriptsize$\pm$0.25}}
    & \makecell{\textbf{65.46}\\{\scriptsize$\pm$0.24}}
    & \makecell{\textbf{61.21}\\{\scriptsize$\pm$0.35}}
    & \makecell{\textbf{62.11}\\{\scriptsize$\pm$0.41}}
    & \makecell{\textbf{61.48}\\{\scriptsize$\pm$0.30}}
    & \makecell{\textbf{60.85}\\{\scriptsize$\pm$0.44}}
    & \makecell{\underline{52.81}\\{\scriptsize$\pm$0.43}} \\
    
    \bottomrule
    \end{tabular}
\end{table*}

%% file: tables/ablation.tex
\begin{table}[t]
    \centering
    \caption{Module-wise ablation study. The agr and con denote the agreement and conflict branches, respectively.}
    \label{tab:ablation}
    \setlength{\tabcolsep}{2pt}
    \resizebox{0.9\columnwidth}{!}{
    \small
    \begin{tabular}{l ccc ccc}
        \toprule
        & \multicolumn{3}{c}{\textbf{MIntRec}} & \multicolumn{3}{c}{\textbf{MIntRec2.0}} \\
        \cmidrule(lr){2-4} \cmidrule(lr){5-7}
        \textbf{Method} & \textbf{Acc} & \textbf{WP} & \textbf{WF1} & \textbf{Acc} & \textbf{WP} & \textbf{WF1} \\
        \midrule

        Qwen2.5-Omni-7B
        & 77.98 & 78.26 & 77.50 & 64.88 & 65.39 & 64.23  \\ 
        \midrule

        w/o Hypergraph
        & 78.34 & 78.59 & 78.26 
        & 64.79 & 64.63 & 64.56  \\
        \midrule

        Only Text
        & 78.74 & 79.08 & 78.70 
        & 63.94 & 63.80 & 63.51  \\
        Only Audio 
        & 77.62 & 78.77 & 77.75 
        & 62.65 & 62.34 & 62.19  \\
        Only Visual 
        & 16.40 & 16.04 & 16.10 
        & 13.85 & 12.04 & 12.54  \\
        \midrule
        Text + Audio
        & 79.96 & 80.30 & 79.92 
        & 65.33 & 64.87 & 64.88  \\
        Text + Visual
        & 77.21 & 77.07 & 76.84 
        & 63.40 & 63.25 & 63.16  \\
        Audio + Visual
        & 71.10 & 71.62 & 70.79 
        & 61.12 & 60.99 & 60.88  \\
        \midrule

        Only Unimodal Level
        & 78.34 & 78.76 & 78.24 
        & 65.15 & 65.20 & 65.05  \\
        Only Bimodal Level
        & 79.64 & 80.00 & 79.48 
        & 63.58 & 64.44 & 63.82  \\
        Only Trimodal Level
        & 79.51 & 79.43 & 79.15 
        & 64.07 & 63.74 & 63.69  \\
        w/o Bimodal Level
        & 78.65 & 78.86 & 78.49 
        & 64.22 & 63.95 & 63.92  \\
        w/o Trimodal Level
        & 79.78 & 79.96 & 79.64 
        & 65.29 & 65.16 & 65.08 \\
        \midrule

        w/o agr Prototype
        & 77.66 & 77.62 & 77.30 
        & 64.27 & 64.11 & 64.01  \\
        w/o con Prototype
        & 79.24 & 79.07 & 78.80 
        & 65.16 & 64.82 & 64.85  \\
        w/o Prototypes
        & 80.40 & 80.52 & 80.22 
        & 64.12 & 63.85 & 63.80  \\
        \midrule

        Only agr - w/o Prototype
        & 79.64 & 79.97 & 79.57 
        & 63.76 & 63.48 & 63.43  \\
        w/o agr
        & 79.73 & 80.13 & 79.67 
        & 64.19 & 64.90 & 64.39  \\

        Only con - w/o Prototype
        & 79.60 & 79.63 & 79.44 
        & 64.09 & 64.16 & 63.92  \\
        w/o con
        & 79.69 & 79.78 & 79.43 
        & 64.91 & 64.61 & 64.58  \\
        \midrule

        \rowcolor{gray!12} \textbf{Full Model}
        & \textbf{80.99} & \textbf{81.23} & \textbf{80.79} 
        & \textbf{65.67} & \textbf{65.56} & \textbf{65.46}  \\

        \bottomrule
    \end{tabular}
    }
\end{table}

%% file: tables/loss_ablation.tex
\begin{table}[t]
    \centering
    \caption{Loss ablation study.}
    \label{tab:loss_ablation}
    \setlength{\tabcolsep}{2pt}

    \resizebox{0.9\columnwidth}{!}{
    \small
    \begin{tabular}{l ccc ccc}
        \toprule
        & \multicolumn{3}{c}{\textbf{MIntRec}} & \multicolumn{3}{c}{\textbf{MIntRec2.0}} \\
        \cmidrule(lr){2-4} \cmidrule(lr){5-7}
        \textbf{Method} & \textbf{Acc} & \textbf{WP} & \textbf{WF1} & \textbf{Acc} & \textbf{WP} & \textbf{WF1}  \\
        \midrule

        w/o Diversity Loss
        & 77.53 & 77.46 & 77.13 
        & 63.24 & 63.18 & 63.04  \\

        w/o Sup. Contrastive Loss
        & 79.96 & 80.05 & 79.71 
        & 65.47 & 65.20 & 65.16  \\

        w/o Hierarchical Loss
        & 78.65 & 79.06 & 78.45 
        & 62.73 & 63.52 & 62.85  \\

        Only Classification Loss 
        & 78.79 & 79.00 & 78.59 
        & 62.60 & 62.61 & 62.41  \\

        \midrule

        \rowcolor{gray!12} \textbf{Full Model}
        & \textbf{80.99} & \textbf{81.23} & \textbf{80.79} 
        & \textbf{65.67} & \textbf{65.56} & \textbf{65.46}  \\
        \bottomrule
    \end{tabular}
    }
\end{table}











%% file: 5_conclusion.tex
\section{Conclusion}\label{sec:conclusion}
This paper presented MACH, a hierarchical prototype-guided hypergraph framework for multimodal intent recognition that explicitly models modality agreement and conflict as complementary semantic signals. By progressively reasoning over unimodal, bimodal, and trimodal interactions, MACH learns discriminative and interpretable representations while effectively capturing cross-modal semantics. Extensive experiments on three benchmark datasets demonstrate consistent improvements over strong multimodal baselines, and comprehensive ablation studies validate the contribution of each proposed component. Future work will investigate incorporating broader conversational context and extending the proposed hierarchical agreement-conflict reasoning framework to other multimodal understanding tasks.

%% file: aaai2027.bib
@inproceedings{cont_leanring_nips_2020,
author = {Khosla, Prannay and Teterwak, Piotr and Wang, Chen and Sarna, Aaron and Tian, Yonglong and Isola, Phillip and Maschinot, Aaron and Liu, Ce and Krishnan, Dilip},
title = {{Supervised contrastive learning}},
year = {2020},
isbn = {9781713829546},
publisher = {Curran Associates Inc.},
address = {Red Hook, NY, USA},
booktitle = {Proceedings of the 34th International Conference on Neural Information Processing Systems},
articleno = {1567},
numpages = {13},
location = {Vancouver, BC, Canada},
series = {NIPS '20}
}

@inproceedings{hypergraph_NN_aaai_2019,
author = {Feng, Yifan and You, Haoxuan and Zhang, Zizhao and Ji, Rongrong and Gao, Yue},
title = {{Hypergraph neural networks}},
year = {2019},
isbn = {978-1-57735-809-1},
publisher = {AAAI Press},
url = {https://doi.org/10.1609/aaai.v33i01.33013558},
doi = {10.1609/aaai.v33i01.33013558},
booktitle = {Proceedings of the Thirty-Third AAAI Conference on Artificial Intelligence and Thirty-First Innovative Applications of Artificial Intelligence Conference and Ninth AAAI Symposium on Educational Advances in Artificial Intelligence},
articleno = {437},
numpages = {8},
location = {Honolulu, Hawaii, USA},
series = {AAAI'19/IAAI'19/EAAI'19}
}

@inproceedings{hier_cvpr_26,
  title={{Evolutionary Multimodal Reasoning via Hierarchical Semantic Representation for Intent Recognition}},
  author={Zhou, Qianrui and Xu, Hua and Gu, Yunjin and Wang, Yifan and Li, Songze and Zhang, Hanlei},
  booktitle={Proceedings of the IEEE/CVF Conference on Computer Vision and Pattern Recognition},
  pages={14979--14989},
  year={2026}
}

@inproceedings{magbert,
  title={Integrating multimodal information in large pretrained transformers},
  author={Rahman, Wasifur and Hasan, Md Kamrul and Lee, Sangwu and Zadeh, AmirAli Bagher and Mao, Chengfeng and Morency, Louis-Philippe and Hoque, Ehsan},
  booktitle={Proceedings of the 58th annual meeting of the association for computational linguistics},
  pages={2359--2369},
  year={2020}
}

@article{huang2025mvcl,
  title={MVCL-DAF++: Enhancing Multimodal Intent Recognition via Prototype-Aware Contrastive Alignment and Coarse-to-Fine Dynamic Attention Fusion},
  author={Huang, Haofeng and Han, Yifei and Zhang, Long and Li, Bin and He, Yangfan},
  journal={arXiv preprint arXiv:2509.17446},
  year={2025}
}

@article{zhao2025deep,
  title={Deep learning approaches for multimodal intent recognition: A survey},
  author={Zhao, Jingwei and Wen, Yuhua and Li, Qifei and Hu, Minchi and Zhou, Yingying and Xue, Jingyao and Wu, Junyang and Gao, Yingming and Wen, Zhengqi and Tao, Jianhua and others},
  journal={arXiv preprint arXiv:2507.22934},
  year={2025}
}

@article{zhu2025survey,
  title={A Survey on Multi-modal Intent Recognition: Recent Advances and New Frontiers},
  author={Zhu, Zhihong and Zhang, Fan and Zhang, Yunyan and Sun, Jinghan and Huang, Zhiqi and Long, Qingqing and Xing, Bowen and Wu, Xian},
  journal={Findings of the Association for Computational Linguistics: EMNLP 2025},
  pages={15223--15236},
  year={2025}
}

@inproceedings{mintrec,
  title={{MIntRec}: A new dataset for multimodal intent recognition},
  author={Zhang, Hanlei and Xu, Hua and Wang, Xin and Zhou, Qianrui and Zhao, Shaojie and Teng, Jiayan},
  booktitle={Proceedings of the 30th ACM international conference on multimedia},
  pages={1688--1697},
  year={2022}
}

@inproceedings{mintrec2,
  title={{MIntRec2.0}: A Large-scale Benchmark Dataset for Multimodal Intent Recognition and Out-of-scope Detection in Conversations},
  author={Zhang, Hanlei and Wang, Xin and Xu, Hua and Zhou, Qianrui and Gao, Kai and Su, Jianhua and Li, Wenrui and Chen, Yanting and others},
  booktitle={The Twelfth International Conference on Learning Representations},
year={2024}
}

@inproceedings{TCLMAP_aaai_2024,
  title={Token-level contrastive learning with modality-aware prompting for multimodal intent recognition},
  author={Zhou, Qianrui and Xu, Hua and Li, Hao and Zhang, Hanlei and Zhang, Xiaohan and Wang, Yifan and Gao, Kai},
  booktitle={Proceedings of the AAAI conference on artificial intelligence},
  volume={38},
  number={15},
  pages={17114--17122},
  year={2024}
}

@ARTICLE{PIP-NET_TMM25,
  author={Wang, Binglu and Yang, Kang and Zhao, Yongqiang and Long, Teng and Li, Xuelong},
  journal={IEEE Transactions on Multimedia}, 
  title={Prototype-Based Intent Perception}, 
  year={2023},
  volume={25},
  number={},
  pages={8308-8319},
  doi={10.1109/TMM.2023.3234817}}

@inproceedings{MVCL_DAF_aaai2025,
author = {Hu, Bo and Zhang, Kai and Zhang, Yanghai and Ye, Yuyang},
title = {Adaptive multimodal fusion: dynamic attention allocation for intent recognition},
year = {2025},
isbn = {978-1-57735-897-8},
publisher = {AAAI Press},
doi = {10.1609/aaai.v39i16.33898},
booktitle = {Proceedings of the Thirty-Ninth AAAI Conference on Artificial Intelligence and Thirty-Seventh Conference on Innovative Applications of Artificial Intelligence and Fifteenth Symposium on Educational Advances in Artificial Intelligence},
articleno = {1925},
numpages = {9},
series = {AAAI'25/IAAI'25/EAAI'25}
}

@inproceedings{meld2019_ACL,
    title = "{MELD}: A Multimodal Multi-Party Dataset for Emotion Recognition in Conversations",
    author = "Poria, Soujanya  and
      Hazarika, Devamanyu  and
      Majumder, Navonil  and
      Naik, Gautam  and
      Cambria, Erik  and
      Mihalcea, Rada",
    editor = "Korhonen, Anna  and
      Traum, David  and
      M{\`a}rquez, Llu{\'i}s",
    booktitle = "Proceedings of the 57th Annual Meeting of the Association for Computational Linguistics",
    month = jul,
    year = "2019",
    address = "Florence, Italy",
    publisher = "Association for Computational Linguistics",
    url = "https://aclanthology.org/P19-1050/",
    doi = "10.18653/v1/P19-1050",
    pages = "527--536"
}

@inproceedings{InMuNet_2024_MM,
author = {Zhu, Zhihong and Cheng, Xuxin and Chen, Zhaorun and Chen, Yuyan and Zhang, Yunyan and Wu, Xian and Zheng, Yefeng and Xing, Bowen},
title = {{InMu-Net: Advancing Multi-modal Intent Detection via Information Bottleneck and Multi-sensory Processing}},
year = {2024},
isbn = {9798400706868},
publisher = {Association for Computing Machinery},
address = {New York, NY, USA},
url = {https://doi.org/10.1145/3664647.3681623},
doi = {10.1145/3664647.3681623},
booktitle = {Proceedings of the 32nd ACM International Conference on Multimedia},
pages = {515–524},
numpages = {10},
location = {Melbourne VIC, Australia},
series = {MM '24}
}

@inproceedings{cdpr_icmr_2026,
  title={{Mitigating Multimodal Inconsistency via Cognitive Dual-Pathway Reasoning for Intent Recognition}},
  author={Wang, Yifan and Wang, Peiwu and Chi, Yunxian and Gou, Zhinan and Gao, Kai},
  booktitle={Proceedings of the 2026 International Conference on Multimedia Retrieval},
  pages={797--806},
  year={2026}
}

@inproceedings{wcfmir_aaai26,
author = {Liu, Yi and Yang, Qimeng and Lu, Lanlan},
title = {{Who should I trust? explicit confidence-focused multimodal intent recognition}},
year = {2026},
isbn = {978-1-57735-906-7},
publisher = {AAAI Press},
url = {https://doi.org/10.1609/aaai.v40i28.39565},
doi = {10.1609/aaai.v40i28.39565},
booktitle = {Proceedings of the Fortieth AAAI Conference on Artificial Intelligence and Thirty-Eighth Conference on Innovative Applications of Artificial Intelligence and Sixteenth Symposium on Educational Advances in Artificial Intelligence},
articleno = {2664},
numpages = {8},
series = {AAAI'26/IAAI'26/EAAI'26}
}

@inproceedings{mult_2019_acl,
    title = "Multimodal Transformer for Unaligned Multimodal Language Sequences",
    author = "Tsai, Yao-Hung Hubert  and
      Bai, Shaojie  and
      Liang, Paul Pu  and
      Kolter, J. Zico  and
      Morency, Louis-Philippe  and
      Salakhutdinov, Ruslan",
    editor = "Korhonen, Anna  and
      Traum, David  and
      M{\`a}rquez, Llu{\'i}s",
    booktitle = "Proceedings of the 57th Annual Meeting of the Association for Computational Linguistics",
    month = jul,
    year = "2019",
    address = "Florence, Italy",
    publisher = "Association for Computational Linguistics",
    url = "https://aclanthology.org/P19-1656/",
    doi = "10.18653/v1/P19-1656",
    pages = "6558--6569",
}

@inproceedings{misa_MM_20,
author = {Hazarika, Devamanyu and Zimmermann, Roger and Poria, Soujanya},
title = {MISA: Modality-Invariant and -Specific Representations for Multimodal Sentiment Analysis},
year = {2020},
isbn = {9781450379885},
publisher = {Association for Computing Machinery},
address = {New York, NY, USA},
url = {https://doi.org/10.1145/3394171.3413678},
doi = {10.1145/3394171.3413678},
booktitle = {Proceedings of the 28th ACM International Conference on Multimedia},
pages = {1122–1131},
numpages = {10},
location = {Seattle, WA, USA},
series = {MM '20}
}

@article{LiPLC25,
  author       = {Zhengyi Li and
                  Junjie Peng and
                  Xuanchao Lin and
                  Zesu Cai},
  title        = {Multimodal intent recognition based on text-guided cross-modal attention},
  journal      = {Appl. Intell.},
  volume       = {55},
  number       = {7},
  pages        = {690},
  year         = {2025},
  url          = {https://doi.org/10.1007/s10489-025-06583-2},
  doi          = {10.1007/S10489-025-06583-2},
  bibsource    = {dblp computer science bibliography, https://dblp.org}
}

@article{GMoE-DCAA,
title = {Graph Mixture of Experts with Differential Cross-Attention Alignment for Multimodal Intent Recognition},
journal = {Knowledge-Based Systems},
volume = {349},
pages = {116392},
year = {2026},
issn = {0950-7051},
doi = {https://doi.org/10.1016/j.knosys.2026.116392},
url = {https://www.sciencedirect.com/science/article/pii/S0950705126011184},
author = {Shilin Sun and Wenbin An and Qidong Liu and Fang Nan and Jiahao Nie and Zhi Zeng and Xian-Sheng Hua and Yaqiang Wu and Feng Tian}
}

@article{H2GF-Net,
  title={Hybrid-Scale Heterogeneous Graph Fusion for Multimodal Intent Recognition},
  author={Liu, Rui and Ma, Yuxuan and Li, Haizhou},
  journal={IEEE Transactions on Affective Computing},
  year={2026},
  publisher={IEEE}
}

@inproceedings{al2025hypermodal,
  title={HyperModal: Dynamic Hypergraph Contrastive Learning for Multi-Modal Representation},
  author={Al Kulaib, Lulwah},
  booktitle={2025 IEEE International Conference on Data Mining (ICDM)},
  pages={1023--1029},
  year={2025},
  organization={IEEE}
}

@inproceedings{saifuddin2025hypergcl,
  title={HyperGCL: multi-modal graph contrastive learning via learnable hypergraph views},
  author={Saifuddin, Khaled Mohammed and Ji, Shihao and Akbas, Esra},
  booktitle={2025 International Joint Conference on Neural Networks (IJCNN)},
  pages={1--8},
  year={2025},
  organization={IEEE}
}

@inproceedings{huang2024sdif,
  title={Sdif-da: A shallow-to-deep interaction framework with data augmentation for multi-modal intent detection},
  author={Huang, Shijue and Qin, Libo and Wang, Bingbing and Tu, Geng and Xu, Ruifeng},
  booktitle={ICASSP 2024-2024 IEEE International Conference on Acoustics, Speech and Signal Processing (ICASSP)},
  pages={10206--10210},
  year={2024},
  organization={IEEE}
}
